\documentclass[nohyperref]{article}

\usepackage{microtype}
\usepackage{caption}
\usepackage{graphicx}
\usepackage{subfigure}
\usepackage{booktabs} 

\usepackage{hyperref}

\usepackage[accepted]{icml2023}

\usepackage{amsmath}
\usepackage{amssymb}
\usepackage{mathtools}
\usepackage{amsthm}
\usepackage{hyperref}
\usepackage{booktabs}
\usepackage[capitalize,noabbrev]{cleveref}

\theoremstyle{plain}

\theoremstyle{definition}

\theoremstyle{remark}

\usepackage[textsize=tiny]{todonotes}

\icmltitlerunning{Magnification Invariant Medical Image Analysis}

\begin{document}

\twocolumn[
\icmltitle{Magnification Invariant Medical Image Analysis: A Comparison of Convolutional Networks, Vision Transformers, and Token Mixers}



\icmlsetsymbol{equal}{*}

\begin{icmlauthorlist}
\icmlauthor{Pranav Jeevan}{yyy}
\icmlauthor{Nikhil Cherian Kurian}{yyy}
\icmlauthor{Amit Sethi}{yyy}
\end{icmlauthorlist}

\icmlaffiliation{yyy}{Department of Electrical Engineering, Indian Institute of Technology Bombay, Mumbai, India}

\icmlcorrespondingauthor{Pranav Jeevan}{194070025@iitb.ac.in}

\icmlkeywords{Machine Learning, ICML}

\vskip 0.3in
]



\printAffiliationsAndNotice{}  

\begin{abstract}
 Convolution Neural Networks (CNNs) are widely used in medical image analysis, but their performance degrade when the  magnification of testing images differ from the training images. The inability of CNNs to generalize across magnification scales can result in sub-optimal performance on external datasets. This study aims to evaluate the robustness of various deep learning architectures in the analysis of breast cancer histopathological images with varying magnification scales at training and testing stages. Here we explore and compare the performance of multiple deep learning architectures, including CNN-based ResNet and MobileNet, self-attention-based Vision Transformers and Swin Transformers, and token-mixing models, such as FNet, ConvMixer, MLP-Mixer, and WaveMix. The experiments are conducted using the BreakHis \cite{spanhol2015dataset} dataset, which contains breast cancer histopathological images at varying magnification levels. We show that performance of WaveMix is invariant to the magnification of training and testing data and can provide stable and good classification accuracy. These evaluations are critical in identifying deep learning architectures that can robustly handle changes in magnification scale, ensuring that scale changes across anatomical structures do not disturb the inference results. 
\end{abstract}

\section{Introduction}
\label{submission}
Computer aided medical image analysis has become a critical component in the diagnosis and treatment of various diseases \cite{chakraborty2023overview,duncan2000medical}. Deep learning models, such as Convolution neural networks (CNNs), have shown exceptional performance in analyzing medical images, including magnetic resonance imaging (MRI), computed tomography (CT), and histology images \cite{chan2020deep}. However, the performance of these models can be affected by several factors, including variations in image quality, lighting conditions, and magnification scales. In particular, changes in magnification scales between training and testing datasets can significantly impact the accuracy and robustness of deep learning models in medical image analysis \cite{gupta2017breast}.



CNNs are cited to be the most commonly used deep learning architecture for medical image analysis \cite{li2014medical}. However CNN, can struggle when it comes to handling medical images with anatomical features at varying magnification scales. In general, training a CNN on images at a specific magnification scale may result in good performance on that scale, but this performance may not generalize well to other magnification scales \cite{alkassar2021going}. This is a significant limitation when analysing medical imaging modalities like histology images where slight to moderate changes in magnification variability is common. The inability of CNN to generalize across magnification scales leads to sub-optimal inference performance on external datasets \cite{gupta2017breast}. Though, augmenting input images with perturbations in scales can slightly improve performance of CNNs, it is also important to explore or develop more robust deep learning architectures that can generate features that are inherently invariant to the changes in scale of input images. Such architectures should be designed to capture the important features in the images, regardless of the shift in the magnification scale, in order to provide robust performance for medical image analysis in a clinical settings. 

In this study, we evaluate the robustness of multiple popular deep learning architectures including CNN based architectures such as ResNet \cite{he2016deep} and MobileNet \cite{Howard2017-rm}, Self-attention based architectures such as Vision Transformers (VIT) \cite{dosovitskiy2021an} and Swin Transformers \cite{liu2021swin}, and token mixing models such as Fourier-Net (FNet) \cite{lee2021fnet}, ConvMixer \cite{trockman2022patches}, Multi-Layer Perceptron-Mixer (MLP-Mixer) \cite{tolstikhin2021mlp}, and WaveMix \cite{https://doi.org/10.48550/arxiv.2205.14375}. Our aim is to compare the performance of these deep learning models when the magnification of the test data differs from the training data. The BreakHis \cite{spanhol2015dataset} dataset , which includes breast cancer histopathological images at varying magnification levels, is utilized for our experiments. The empirical performance differences between the deep learning models will be used to determine the most robust architecture for histopathological image analysis.

\section{Experiments}
\subsection{Dataset}
We utilize the BreakHis \cite{spanhol2015dataset} dataset, which is a well-known public dataset in the field of digital breast histopathology for our experiments. It has been widely used in the development and evaluation of computer-aided diagnosis (CAD) systems for breast cancer diagnosis. It provides a challenging benchmark for the development of CAD systems due to the inherent large variations in tissue appearances.

The dataset consist of  7,909 microscopy images of breast tissue biopsy specimens from 82 patients diagnosed with either benign or malignant breast tumors. The images are collected from four different institutions and are of four different magnifications scales - 40X, 100X, 200X and 400X.

In addition to the malignancy information of each image, the dataset is further annotated with information like the patient's age, the sub-type of malignancy and the type of biopsy. The dataset is slightly imbalanced in terms of the distribution of benign and malignant cases and the distribution of different magnifications. In the dataset there are 5,429 malignant cases whereas benign cases are only about 2,480. 

As the BreakHis \cite{spanhol2015dataset} dataset contains multiple images at different magnification levels, the dataset serves as a challenging and representative testbed for evaluating the robustness of deep learning architectures across the different magnification levels or scales. These evaluations will be carried out by training some of the recently reported deep learning architecture on one magnification level of the BreakHis \cite{spanhol2015dataset} dataset and testing these trained models across multiple held-out magnification levels. Observing the  average test accuracy   on the different  magnification levels can hence reveal the robustness of deep learning architectures to varying image magnification at inference..


\subsection{Models}

\begin{figure*}[ht]
\vskip 0.2in
\begin{center}
\includegraphics[scale=0.75]{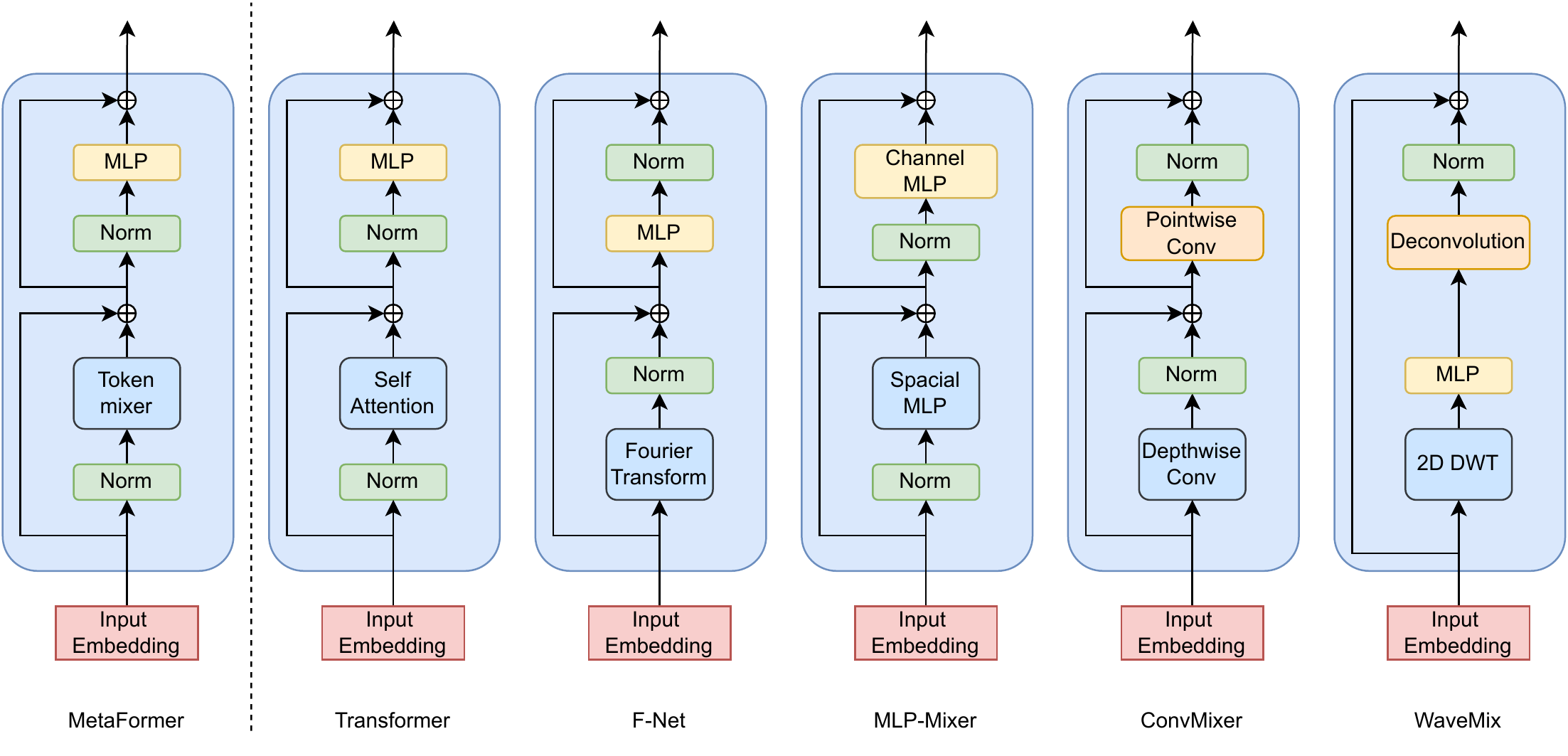}
\caption{Architectures of various token-mixers along with the general MetaFormer block. The token-mixing operation in different models is performed by different operations, such as spacial MLP, depth-wise convolution, self-attention, Fourier and Wavelet transforms}
\label{fig:token}
\end{center}
\vskip -0.2in
\end{figure*}

For CNN based models, we compared performance using ResNet-18, ResNet-34 and ResNet-50 from the ResNet family \cite{he2016deep}, and MobileNetV3-small-0.50, MobileNetV3-small-0.75 and MobileNetV3-small-100 from MobileNet family of models. We used ViT-Tiny, ViT-Small and ViT-Base (all using patch size of 16, see \cite{dosovitskiy2021an}) along with Swin-Tiny and Swin-Base (all using patch size of 4 and window size of 7, see \cite{liu2021swin}) for the experiments.

\subsubsection{Token-Mixers}
Token-mixers are the family of models which uses an architecture similar to MetaFormer \cite{yu2022metaformer} as its fundamental block as shown in ~\Cref{fig:token}. Transformer models can be considered as token-mixing model which uses self-attention for token-mixing. Other token-mixers use Fourier transforms (FNet) \cite{lee2021fnet}, Wavelet transforms (WaveMix) \cite{https://doi.org/10.48550/arxiv.2205.14375}, spatial-MLP (MLP-Mixer) \cite{tolstikhin2021mlp} or depth-wise convolutions (ConvMixer) \cite{trockman2022patches} for token-mixing. Token-mixing models have been shown to be more efficient in terms of parameters and computation compared to attention-based transformers \cite{yu2022metaformer}.

FNet \cite{lee2021fnet} was actually designed for natural language processing (NLP) tasks and was designed to handle 1D inputs sequences. We have used the 2D-FNet, i.e., a modified FNet that used a 2D Fourier transform for spacial token-mixing instead of a 1D Fourier transform used in FNet. The 2-D FNet can process images in the 2D form without the need to unroll it into sequence of patches or pixels as done in transformer and FNet. We experimented by varying the embedding dimension and number of layers to get the best model.

WaveMix \cite{https://doi.org/10.48550/arxiv.2205.14375} uses 2D-Discrete Wavelet transform (2D-DWT) for token-mixing. We experimented by varying the embedding dimension, number of layers and number of levels of 2D-DWT used in WaveMix to get the model which gives highest validation accuracy in the dataset. 

ConvMixer \cite{trockman2022patches} uses depth-wise convolution for spacial token-mixing and point-wise convolutions for channel toke-mixing. We used ConvMixer-1536/20, ConvMixer-768/32, and ConvMixer-1024/20 available in Timm model library for our experiments. 

MLP-Mixer \cite{tolstikhin2021mlp} uses spacial MLP and channel MLP to mix tokens. We used MLP-Mixer-Small (patch size of 16) and MLP-Mixer-Base (patch size of 16) in our experiments.

\subsection{Implementation details}

The dataset was divided into train, validation and test sets in the ratio 7:1:2 for each of the magnification. Due to limited computational resources, the maximum number of training epochs was set to 300. All experiments were done with a single 80 GB Nvidia A100 GPU. \emph{No pre-trained weights was used for any of the models}. We used the ResNet, MobileNet, Vision transformer, Swin transformer, ConvMixer and MLP-Mixer available in Timm (PyTorch Image Models) library~\cite{rw2019timm}\footnote{available at \url{http://github.com/rwightman/pytorch-image-models/}} Since WaveMix and FNet  was unavailable in Timm library, it was implemented from original paper. The Timm training script~\cite{rw2019timm} with default hyper-parameter values was used to train all the models. Cross-entropy loss was used for training. We used automatic mixed precision in PyTorch during training to optimize speed and memory consumption. 

The images were resized to $672\times448$ for the experiments. Transformer-based models and MLP-Mixer required the images to be resized to certain specific sizes like $224\times224$ or $384\times384$. We trained models of varying sizes belonging to the same architecture on the training set and evaluated it on validation set to find the model size that gives the best performance on the Breakhis \cite{spanhol2015dataset} dataset. The model size with highest average validation performance over all magnifications was used for evaluation using test set.

The maximum batch-size was set to 128. For larger models, we reduced the batch-size so that it can fit in the GPU. Top-1 accuracy on the test set of the best of three runs with random initialization is reported as a generalization metric based on prevailing protocols~\cite{hassani2021escaping}. 

\begin{figure}[ht]
\vskip 0.2in
\begin{center}
\includegraphics[scale=0.68]{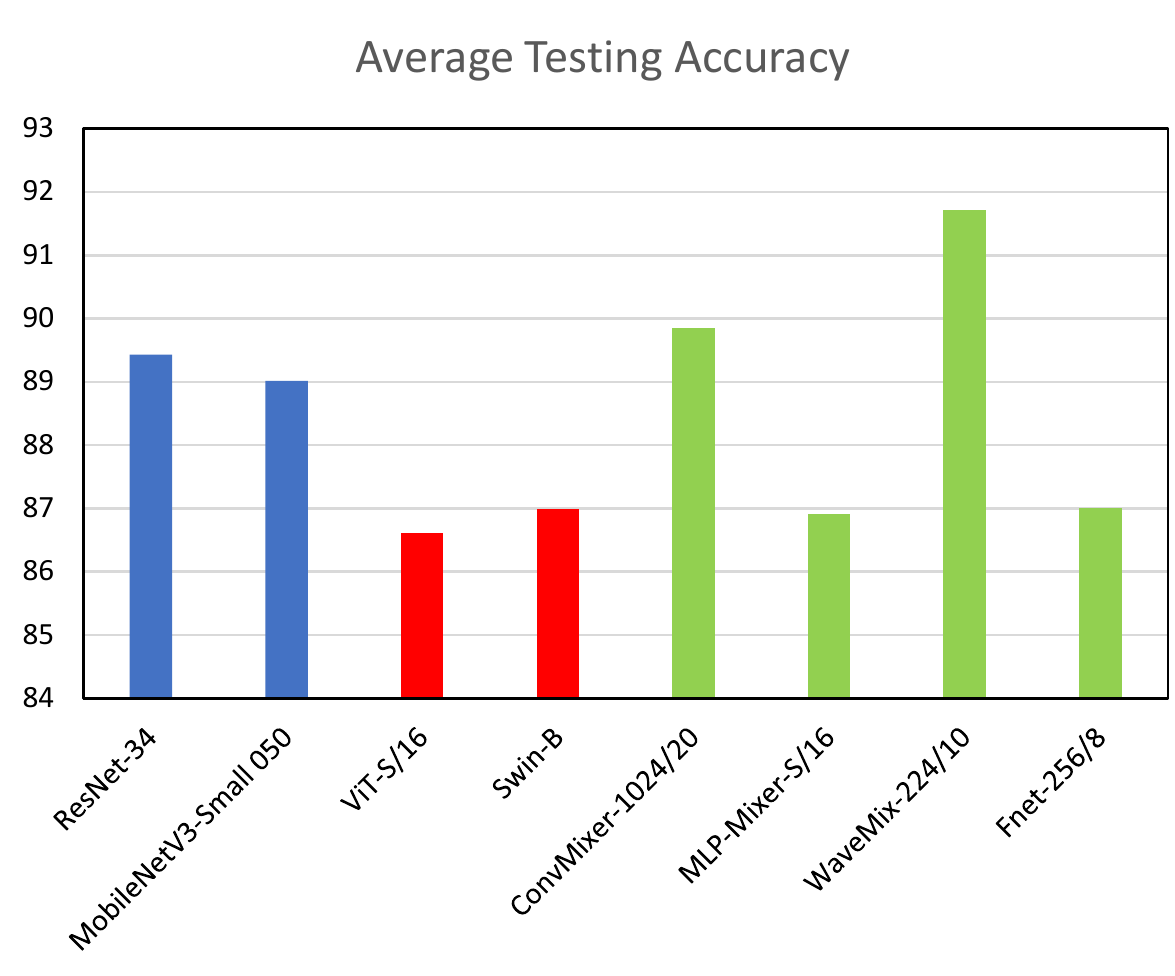}
\caption{Average of all test accuracy reported for various training magnifications for each of the models compared}
\label{fig:test}
\end{center}
\vskip -0.2in
\end{figure}

\section{Results and Discussions}

\
\begin{table*}[]
\centering
\caption{Results of Inter-magnification classification performance of all CNN, transformers and token-mixers on Breakhis \cite{spanhol2015dataset} dataset. Accuracy on test set is reported.}
\vspace{0.5 cm}
\centering

\includegraphics[scale=0.70]{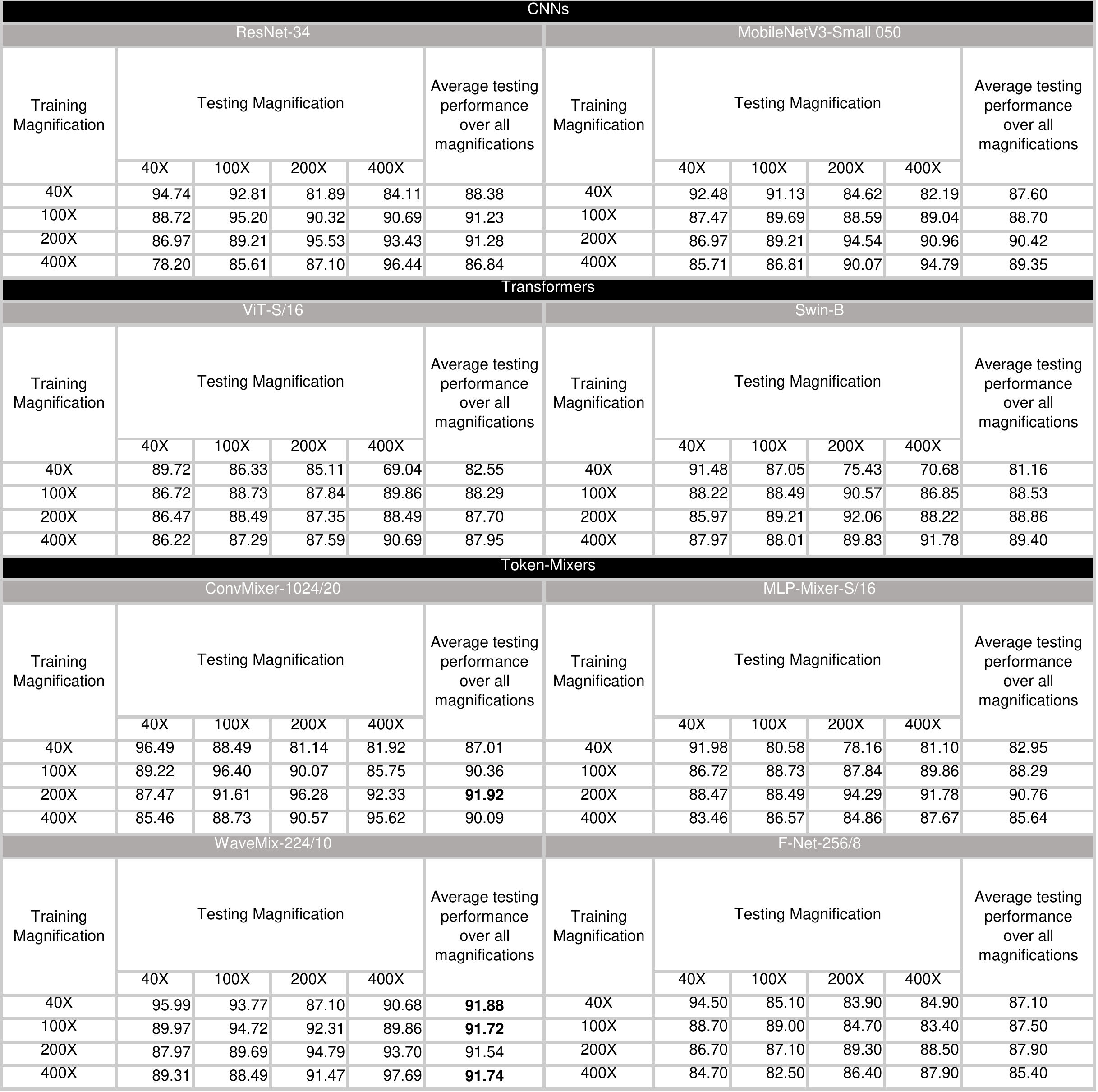}
\label{tab:result}
\end{table*}

\begin{figure}[ht]
\vskip 0.2in
\begin{center}
\includegraphics[scale=0.83]{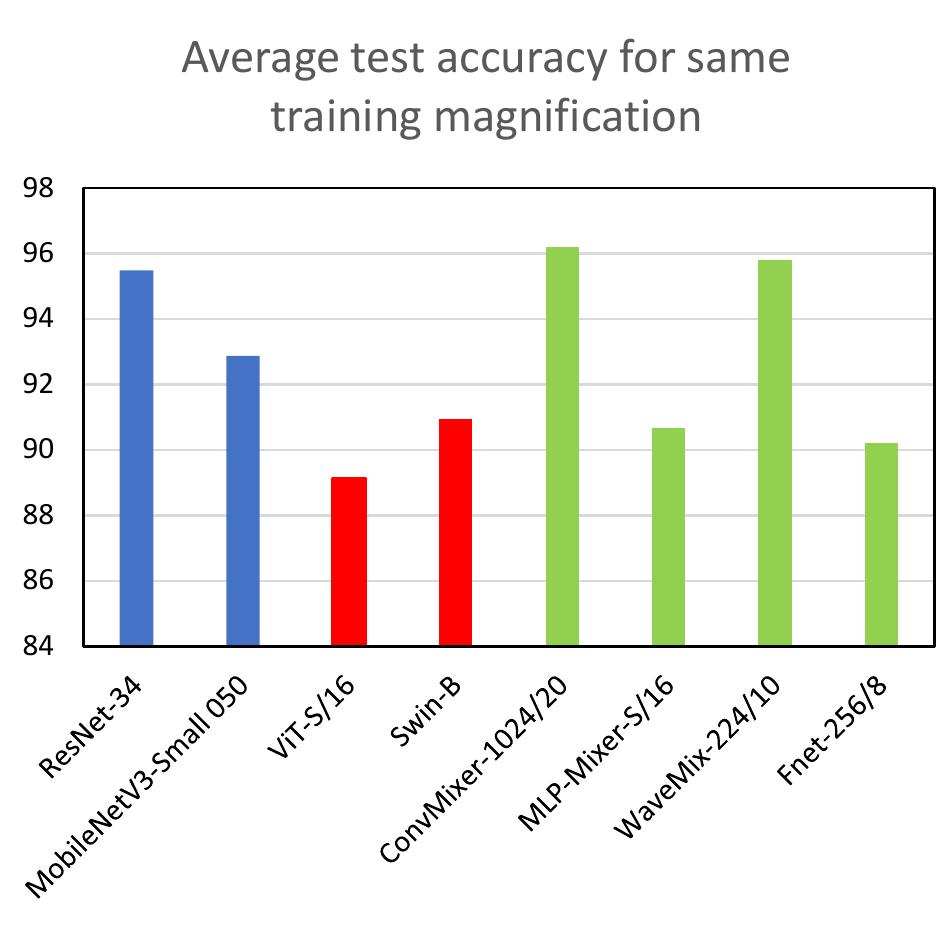}
\caption{Average of test accuracy when training and testing was done on same magnification for each model is compared}
\label{fig:diag}
\end{center}
\vskip -0.2in
\end{figure}

The Inter-magnification classification performance of all the best performing model variants of CNN, transformer and token-mixer models are shown in ~\Cref{tab:result}. We can see that WaveMix performs better than all the other models in maintaining high performance across different testing magnifications. Only ConvMixer, another token-mixer could perform better than WaveMix in one magnification ($200\times$). We also observe that the accuracy of WaveMix  is the most stable, never falling bellow 87\%. Other models that perform well, such as, ConvMixer  and ResNet-34, suffers from unstable performance with accuracy falling to 81\% and 78\% respectively. The better performance of WaveMix is due to the ability of 2D wavelet transform to efficiently mix token information and the subsequent use of deconvolution layers which also aids in rapid expansion of receptive field after each wavelet block.

We also see from~\Cref{fig:test} that WaveMix performs the best amoung all models when we take the overall average of all the average testing accuracy over all magnifications. We observe that the performance of token-mixers (green) like MLP-Mixer and FNet is comparable to that of transformer based models (red). CNN-based models (blue) perform better than transformer models. 

\Cref{fig:diag} show the average of test accuracy when training and testing was done on same magnifications. We observe that ConvMixer performs better than WaveMix when train and test magnifications are same. Even ResNet-34 is performing almost on par with WaveMix and ConvMixer. This shows that even though other models perform well when magnification of training and test data are same, they cannot translate that performance when magnification of training and testing set differs from each other. WaveMix is mostly invariant to this change of magnification between train and test data and is able to provide consistent performance compared to other CNN, transformer and token-mixing models.

FNet consumed largest GPU RAM (4-8$\times$ more) compared to other architectures. CNN-based models perform much better than transformer model-based models in Breakhis classification. There is a significant drop in performance when the transformer-based models are trained on 40$\times$ magnification and tested. Similar drop in accuracy for 40$\times$ magnification training was observed for MLP-Mixer.

\section{Conclusions}
In conclusion, our study evaluated the robustness of various deep learning models for histopathological image analysis under different testing magnifications. We compared ResNet, MobileNet, Vision Transformers, Swin Transformers, Fourier-Net, ConvMixer, MLP-Mixer, and WaveMix using the BreakHis \cite{spanhol2015dataset} dataset. Our experiments demonstrated that the WaveMix architecture, which intrinsically incorporates multi-resolution features, is the most robust model to changes in inference magnification. We observed a stable accuracy of at least 87\% across all test scenarios. These findings highlight the importance of implementing a robust architecture, such as WaveMix, not only for histopathological image analysis but also for medical image analysis in general. This would help to ensure that anatomical features of diverse scales do not influence the accuracy of deep learning-based systems, thereby improving the reliability of diagnostic inference in clinical practice.
\nocite{langley00}

\bibliography{example_paper}

\begin{thebibliography}{19}
\providecommand{\natexlab}[1]{#1}
\providecommand{\url}[1]{\texttt{#1}}
\expandafter\ifx\csname urlstyle\endcsname\relax
  \providecommand{\doi}[1]{doi: #1}\else
  \providecommand{\doi}{doi: \begingroup \urlstyle{rm}\Url}\fi

\bibitem[Alkassar et~al.(2021)Alkassar, Jebur, Abdullah, Al-Khalidy, and
  Chambers]{alkassar2021going}
Alkassar, S., Jebur, B.~A., Abdullah, M.~A., Al-Khalidy, J.~H., and Chambers,
  J.~A.
\newblock Going deeper: magnification-invariant approach for breast cancer
  classification using histopathological images.
\newblock \emph{IET Computer Vision}, 15\penalty0 (2):\penalty0 151--164, 2021.

\bibitem[Chakraborty \& Mali(2023)Chakraborty and
  Mali]{chakraborty2023overview}
Chakraborty, S. and Mali, K.
\newblock An overview of biomedical image analysis from the deep learning
  perspective.
\newblock \emph{Research Anthology on Improving Medical Imaging Techniques for
  Analysis and Intervention}, pp.\  43--59, 2023.

\bibitem[Chan et~al.(2020)Chan, Samala, Hadjiiski, and Zhou]{chan2020deep}
Chan, H.-P., Samala, R.~K., Hadjiiski, L.~M., and Zhou, C.
\newblock Deep learning in medical image analysis.
\newblock \emph{Deep Learning in Medical Image Analysis: Challenges and
  Applications}, pp.\  3--21, 2020.

\bibitem[Dosovitskiy et~al.(2021)Dosovitskiy, Beyer, Kolesnikov, Weissenborn,
  Zhai, Unterthiner, Dehghani, Minderer, Heigold, Gelly, Uszkoreit, and
  Houlsby]{dosovitskiy2021an}
Dosovitskiy, A., Beyer, L., Kolesnikov, A., Weissenborn, D., Zhai, X.,
  Unterthiner, T., Dehghani, M., Minderer, M., Heigold, G., Gelly, S.,
  Uszkoreit, J., and Houlsby, N.
\newblock An image is worth 16x16 words: Transformers for image recognition at
  scale.
\newblock In \emph{International Conference on Learning Representations}, 2021.
\newblock URL \url{https://openreview.net/forum?id=YicbFdNTTy}.

\bibitem[Duncan \& Ayache(2000)Duncan and Ayache]{duncan2000medical}
Duncan, J.~S. and Ayache, N.
\newblock Medical image analysis: Progress over two decades and the challenges
  ahead.
\newblock \emph{IEEE transactions on pattern analysis and machine
  intelligence}, 22\penalty0 (1):\penalty0 85--106, 2000.

\bibitem[Gupta \& Bhavsar(2017)Gupta and Bhavsar]{gupta2017breast}
Gupta, V. and Bhavsar, A.
\newblock Breast cancer histopathological image classification: is
  magnification important?
\newblock In \emph{Proceedings of the IEEE conference on computer vision and
  pattern recognition workshops}, pp.\  17--24, 2017.

\bibitem[Hassani et~al.(2021)Hassani, Walton, Shah, Abuduweili, Li, and
  Shi]{hassani2021escaping}
Hassani, A., Walton, S., Shah, N., Abuduweili, A., Li, J., and Shi, H.
\newblock Escaping the big data paradigm with compact transformers, 2021.

\bibitem[He et~al.(2016)He, Zhang, Ren, and Sun]{he2016deep}
He, K., Zhang, X., Ren, S., and Sun, J.
\newblock Deep residual learning for image recognition.
\newblock In \emph{Proceedings of the IEEE conference on computer vision and
  pattern recognition}, pp.\  770--778, 2016.

\bibitem[Howard et~al.(2017)Howard, Zhu, Chen, Kalenichenko, Wang, Weyand,
  Andreetto, and Adam]{Howard2017-rm}
Howard, A.~G., Zhu, M., Chen, B., Kalenichenko, D., Wang, W., Weyand, T.,
  Andreetto, M., and Adam, H.
\newblock {MobileNets}: Efficient convolutional neural networks for mobile
  vision applications.
\newblock April 2017.

\bibitem[Jeevan et~al.(2022)Jeevan, Viswanathan, S, and
  Sethi]{https://doi.org/10.48550/arxiv.2205.14375}
Jeevan, P., Viswanathan, K., S, A.~A., and Sethi, A.
\newblock Wavemix: A resource-efficient neural network for image analysis,
  2022.
\newblock URL \url{https://arxiv.org/abs/2205.14375}.

\bibitem[Langley(2000)]{langley00}
Langley, P.
\newblock Crafting papers on machine learning.
\newblock In Langley, P. (ed.), \emph{Proceedings of the 17th International
  Conference on Machine Learning (ICML 2000)}, pp.\  1207--1216, Stanford, CA,
  2000. Morgan Kaufmann.

\bibitem[Lee-Thorp et~al.(2021)Lee-Thorp, Ainslie, Eckstein, and
  Ontanon]{lee2021fnet}
Lee-Thorp, J., Ainslie, J., Eckstein, I., and Ontanon, S.
\newblock Fnet: Mixing tokens with fourier transforms.
\newblock \emph{arXiv preprint arXiv:2105.03824}, 2021.

\bibitem[Li et~al.(2014)Li, Cai, Wang, Zhou, Feng, and Chen]{li2014medical}
Li, Q., Cai, W., Wang, X., Zhou, Y., Feng, D.~D., and Chen, M.
\newblock Medical image classification with convolutional neural network.
\newblock In \emph{2014 13th international conference on control automation
  robotics \& vision (ICARCV)}, pp.\  844--848. IEEE, 2014.

\bibitem[Liu et~al.(2021)Liu, Lin, Cao, Hu, Wei, Zhang, Lin, and
  Guo]{liu2021swin}
Liu, Z., Lin, Y., Cao, Y., Hu, H., Wei, Y., Zhang, Z., Lin, S., and Guo, B.
\newblock Swin transformer: Hierarchical vision transformer using shifted
  windows.
\newblock In \emph{Proceedings of the IEEE/CVF international conference on
  computer vision}, pp.\  10012--10022, 2021.

\bibitem[Spanhol et~al.(2015)Spanhol, Oliveira, Petitjean, and
  Heutte]{spanhol2015dataset}
Spanhol, F.~A., Oliveira, L.~S., Petitjean, C., and Heutte, L.
\newblock A dataset for breast cancer histopathological image classification.
\newblock \emph{Ieee transactions on biomedical engineering}, 63\penalty0
  (7):\penalty0 1455--1462, 2015.

\bibitem[Tolstikhin et~al.(2021)Tolstikhin, Houlsby, Kolesnikov, Beyer, Zhai,
  Unterthiner, Yung, Steiner, Keysers, Uszkoreit, et~al.]{tolstikhin2021mlp}
Tolstikhin, I.~O., Houlsby, N., Kolesnikov, A., Beyer, L., Zhai, X.,
  Unterthiner, T., Yung, J., Steiner, A., Keysers, D., Uszkoreit, J., et~al.
\newblock Mlp-mixer: An all-mlp architecture for vision.
\newblock \emph{Advances in neural information processing systems},
  34:\penalty0 24261--24272, 2021.

\bibitem[Trockman \& Kolter(2022)Trockman and Kolter]{trockman2022patches}
Trockman, A. and Kolter, J.~Z.
\newblock Patches are all you need?
\newblock \emph{arXiv preprint arXiv:2201.09792}, 2022.

\bibitem[Wightman(2019)]{rw2019timm}
Wightman, R.
\newblock Pytorch image models.
\newblock \url{https://github.com/rwightman/pytorch-image-models}, 2019.

\bibitem[Yu et~al.(2022)Yu, Luo, Zhou, Si, Zhou, Wang, Feng, and
  Yan]{yu2022metaformer}
Yu, W., Luo, M., Zhou, P., Si, C., Zhou, Y., Wang, X., Feng, J., and Yan, S.
\newblock Metaformer is actually what you need for vision.
\newblock In \emph{Proceedings of the IEEE/CVF conference on computer vision
  and pattern recognition}, pp.\  10819--10829, 2022.

\end{thebibliography}
\bibliographystyle{icml2023}



\end{document}